# Free-space Realization of Tunable Pin-like Optical Vortex Beams

DOMENICO BONGIOVANNI[1,2,7], DENGHUI LI[1,7], MIHALIS GOUTSOULAS[3], HAO WU[1], YI HU[1], DAOHONG SONG[1], ROBERTO MORANDOTTI[1,4,8], NIKOLAOS K. EFREMIDIS[1,3,5], AND ZHIGANG CHEN[1,6,9]

[1]*MOE Key Laboratory of Weak-Light Nonlinear Photonics, TEDA Applied Physics Institute and School of Physics, Nankai University, Tianjin 300457, China*
[2]*INRS-EMT, 1650 Blvd. Lionel-Boulet, Varennes, QC J3X 1S2, Canada*
[3]*Department of Applied Mathematics, University of Crete, Heraklion, Crete 71409, Greece*
[4]*Institute of Fundamental and Frontier Sciences, University of Electronic Science and Technology of China, Chengdu 610054, China*
[5]*Institute of Applied and Computational Mathematics, FORTH, Heraklion, Crete 70013, Greece*
[6]*Department of Physics & Astronomy, San Francisco State University, San Francisco, CA 94132, USA*
[7]*These authors contributed equally to this work*
[8]*morandotti@emt.inrs.ca*
[9]*zgchen@nankai.edu.cn*

**Abstract:** We demonstrate, both analytically and experimentally, free-space pin-like optical vortex beams (POVBs). Such angular-momentum-carrying beams feature tunable peak intensity and undergo robust anti-diffracting propagation, realized by judiciously modulating both the amplitude and the phase profile of a standard laser beam. Specifically, they are generated by superimposing a radially-symmetric power-law phase on a helical phase structure, which allows the inclusion of an orbital angular momentum term to the POVBs. During propagation in free-space, these POVBs initially exhibit autofocusing dynamics, and subsequently, their amplitude patterns morph into a high-order Bessel-like profile characterized by a hollow-core and an annular main-lobe with a constant or tunable width during propagation. In contrast with numerous previous endeavors on Bessel beams, our work represents the first demonstration of long-distance free-space generation of optical vortex "pins", with their peak intensity evolution controlled by the impressed amplitude structure. Both the Poynting vectors and the optical radiation forces associated with these beams are also numerically analyzed, revealing novel properties that may be useful for a wide range of applications.

## 1. Introduction

During the last decades, there has been an increasing interest in the study of structured light and associated optical beam shaping techniques [1, 2]. Prior to this enthusiasm, optical vortex beams constituting the fundamental element of singular optics [3-5] have already attracted a great deal of attention due to the peculiar characteristics associated with their phase singularity, topological charge, and hollow intensity distribution [6-12]. Their ability to carry orbital angular momentum (OAM), transferrable to an illuminated object, along with a number of related properties has further paved the way towards novel opportunities for scientific research and advanced applications [7]. These include, for example, optical tweezers and spanners [8, 9] and high-order quantum entanglement [10, 11]. The hollow-core intensity shape consents to circumvent the particles being susceptible to elevated absorptive heating, due to the high-intensity peak located at the center of the fundamental Gaussian beam typically emitted by a standard laser. All these applications require a comprehensive understanding of the intensity distribution and OAM flux within optical vortices. In this regard, a larger variety of optical vortex structures has been reported in the literature [6, 7]; some classical examples are optical Laguerre-Gaussian [13-15], high-order Bessel [16, 17], Helical-Ince-Gaussian [18, 19], and vector vortex [20, 21] beams. Among them, high-order Bessel beams offer significant advantages with respect to the other hollow beams because of their nondiffracting properties, which make them ideal for long-distance particle and atom transport [22]. It has also been shown that it is possible to generate higher-order Bessel-like beams with radius of the hollow core and maximum intensity fully controllable as a function of the propagation distance [23]. In general, diffraction is an undesired effect of light that causes beam expansion and peak intensity deterioration. A great amount of research efforts has been devoted in recent years to find optical beams that are capable to counteract diffraction in different long-distance frameworks [24-27]. Although nondiffracting beams can be achieved via nonlinear effects forming spatial solitons [28-30] or via photonic structures [31], the requirement of nonlinear media and/or complex structure design could hamper the applicability of such beams. In free space, the first introduction of optical Bessel beams [32, 33] has stimulated the generation of a multitude of light beams with propagation-invariant properties such as Mathieu [34, 35] and Bessel-like [23, 36-38] beams.

Since the Airy wave packet was introduced to optics [2, 39, 40], the nondiffracting properties of the light have also been extended beyond a straight-line propagation into curved paths, where various families of self-accelerating beams that are mostly based on an Airy-like profile have been proposed and demonstrated [41, 42]. Of particular interest is the class of so-called "abruptly autofocusing beams", especially for potential uses in material ablation and medical surgery, driven by the fact that such radially-symmetric self-accelerating beams can exhibit a low-intensity propagation combined with a controllable abrupt autofocusing right before a target [43-47]. Further research advances have also proposed the propagation of autofocusing vortex beams carrying OAM [48]. Due to their intrinsic characteristics, all these optical beams present significant diffraction and hence a rapid decrease in their peak intensities during subsequent propagation after the focal point. Applications exploiting the intrinsic proprieties of nondiffractive beams are benefited in many areas, including, biomedical imaging, filamentation, particle manipulation, and free-space optical communications [2, 49-55]. Ideally, every nondiffractive beam carries infinite power content, thanks to which it can counteract diffraction indefinitely. Such a "perfect generation" with a stable amplitude profile over any propagation distance would require the illumination of an infinite transforming element with an ideal plane wave. Nevertheless, due to physical limitations present in all experimental settings, real beams must be truncated by an aperture and, as a consequence, diffraction takes place, resulting in anti-diffracting beams with propagation distances extending from only a few centimeters to meters. (Note that the term "anti-diffracting beam" is now widely used to refer to an arbitrary structured light that is capable to counteract diffraction for a certain range of distances, although it is not "diffraction-free" in a strict sense. Such a mixed-use of terms can be confusing because anti-diffracting could be associated with a special propagation regime [56], such as free-scale dynamics achieved in a particular type of nonlinear media [57-59].)

Recently, a new class of anti-diffracting light waves, named "optical pin beams" (OPBs), has been demonstrated, showing a robust propagation through atmosphere turbulence over distances of kilometers when compared to a standard Gaussian beam [27]. Further research advances have also generalized the class of OPBs and demonstrated their robustness and intensity stability even in a strong scattering medium [60], with a better performance when compared with other nondiffracting beams such as abruptly autofocusing [43, 45, 46] and Bessel [32] type beams.

In this paper, we report the first experimental demonstration of long-distance generation of pin-like optical vortex beams (POVBs) in free-space, representing a topological extension of the previously-introduced OPBs [27, 60]. We analytically analyze the existence of such POVBs. Compared to previous works based on the fundamental Bessel configurations, the POVBs exhibit an initial autofocusing stage followed by an amplitude reshaping into a high-order Bessel-like beam, but the hollow-core radius and the annular main-lobe width vary with the propagation distance. Interestingly, the peak intensity evolution can be easily controlled by imposing a properly-designed amplitude modulation. The POVBs are investigated not only from the viewpoint of their propagation properties, where several examples are numerically and experimentally illustrated but also from their energy characteristics through calculating the associated Poynting vectors and optical forces.

## 2. Theory

The analysis starts by introducing the vector potential of an arbitrary optical beam, $\vec{U}(r,\theta,z,t) = \psi(r,\theta,z)\exp(ikz-\omega t)\,\hat{x}$, assuming linear polarization along the $x$-axis directed by the unity vector $\hat{x}$. Under slowly-varying envelope approximation, the linear wave equation describing the beam propagation dynamics can be expressed in cylindrical coordinates as

$$i\frac{\partial \psi}{\partial z} + \frac{1}{2k}\left(\frac{\partial^2 \psi}{\partial r^2} + \frac{1}{r}\frac{\partial \psi}{\partial r} + \frac{1}{r^2}\frac{\partial^2 \psi}{\partial \theta^2}\right) = 0. \tag{1}$$

In Eq. (1), $\psi(r, \theta, z)$ denotes the electric field envelope, where $r$ and $\theta$ represent the radial and azimuthal coordinates, and $z$ is the longitudinal distance. In addition, $k = 2\pi/\lambda$ is the wavenumber of the optical wave, $\lambda$ is the wavelength, $\omega$ is the angular frequency, and $t$ is the time. The integral representation of Eq. (1) is given by the Fresnel integral

$$\psi(r,\theta,z) = -\frac{ik}{2\pi z}\int_0^r\int_0^{2\pi} A(\rho)\exp[i\phi(\rho,\varphi)]\,e^{ik\frac{r^2+\rho^2-2\rho r\cos(\varphi-\vartheta)}{2z}}\rho\,d\rho, \tag{2}$$

where $A(\rho)$, and $\phi(\rho,\varphi) = -kC_\rho(\rho/w_\rho)^\gamma - l\varphi$ are the amplitude and the phase of the optical wave on the input plane, with $\rho$ and $\varphi$ indicating the radial and azimuthal coordinates at the onset of propagation, $C_\rho$ is an arbitrary scaling parameter for the phase, $\gamma$ the power exponent of phase modulation, $l$ the topological charge, and $w_\rho$ the phase normalization factor, respectively. To find an asymptotic solution to Eq. (2), we apply the stationary phase method by carrying out a procedure similar to the one in Ref. [60]. The beam envelope near the symmetry axis can be well-approximated by the following expression

$$\psi(r,z) = i^{-|l|}\sqrt{\frac{2\pi k}{2-\gamma}}A(\rho(z))J_l\left(kr\left(C\gamma z^{\gamma-1}\right)^{\frac{1}{2-\gamma}}\right)\left(C\gamma z^{\frac{\gamma}{2}}\right)^{\frac{1}{2-\gamma}} \times$$
$$\exp\left[i\left(\frac{kr^2}{2z} + \left(C^2\gamma^2 z^\gamma\right)^{\frac{1}{2-\gamma}}\frac{k}{2}\left(1-\frac{2}{\gamma}\right) - \frac{\pi}{4}(1+2|l|) - l\theta\right)\right], \tag{3}$$

where $J_l$ is the $l^{\text{th}}$-order Bessel function, and $\rho(z) = (C\gamma z)^{1/(2-\gamma)}$, with $C$ equal to $C_\rho/w_\rho^\gamma$. The overall propagation range depends on the maximum radius $\rho_m$ at the onset distance through the relation $z_m = \rho_m^{2-\gamma}/(C\gamma)$. In this work, we limit the value of $\gamma$ in the interval between 0 and 2. Additionally, Eq. (3) shows that the peak intensity evolution of a POVB can be controlled by properly engineering the initial amplitude structure $A(\rho(z))$.

## 3. Experimental demonstration

We have experimentally observed the POVBs described theoretically in the previous section. In our experimental setup, illustrated in Fig. 1(a), a phase-only spatial light modulator (SLM), i.e. a liquid crystal device (Holoeye PLUTO-VIS-016, 8-bit grey phase levels, 1920 × 1080 pixels, 8 × 8μm$^2$ pixel area), is employed to modulate the phase of a CW solid-state green laser (MGL-F-532 at $\lambda = 532$nm, waist = 2mm). Before impinging on the surface of the SLM, the Gaussian-like profile of the laser is expanded by a microscope system (40×, NA=0.65, and $f = 300$mm) to illuminate the active area of the device. According to Eq. (2), the generation of a POVB requires to simultaneously reshape both the amplitude and phase profiles of an initial beam in real-space. Since our SLM device does not provide the functionality of direct amplitude modulation, we employ an indirect technique as developed by Davis and his collaborators [61], which consists of encoding both the amplitude and phase information of a target beam onto a phase-only filter and then performing the generation in the Fourier domain. To this purpose, a spherical lens ($f = 300$mm) is used to compute the Fourier transform of the light reflected by the

SLM, and an opaque mask of radius 0.5 mm is also placed at the Fourier plane to block the zero-order diffraction. Increasing the efficiency of the light diffracted to the first diffraction order is important for getting better results. To achieve higher diffraction efficiency, the spectral amplitude encoded into the phase-only mask of the SLM is not the one obtained by directly applying the formula of the encoding process, but it is created by means of a home-made lookup table compensating the eventual amplitude distortions. For detection, a costumed imaging system composed of a spherical lens ($f$ = 100 mm) and a CCD camera (Coherent LaserCam-HR II, 12-bit dynamic range, $1280 \times 1024$ pixels, $4.6 \times 4.6\,\mu m^2$ pixel area) is used to record the transverse intensity distributions of POVBs. Furthermore, to retrieve the corresponding longitudinal intensity evolutions, the imaging system is also mounted on a manual translation stage that allows for recording the beam profiles at selected distances with 1mm-long steps of resolution over long propagation (50 cm) in free-space. We point out that such POVBs can in principle propagate to much longer distances whenever they are not limited by the size of SLM and the lab space [27].

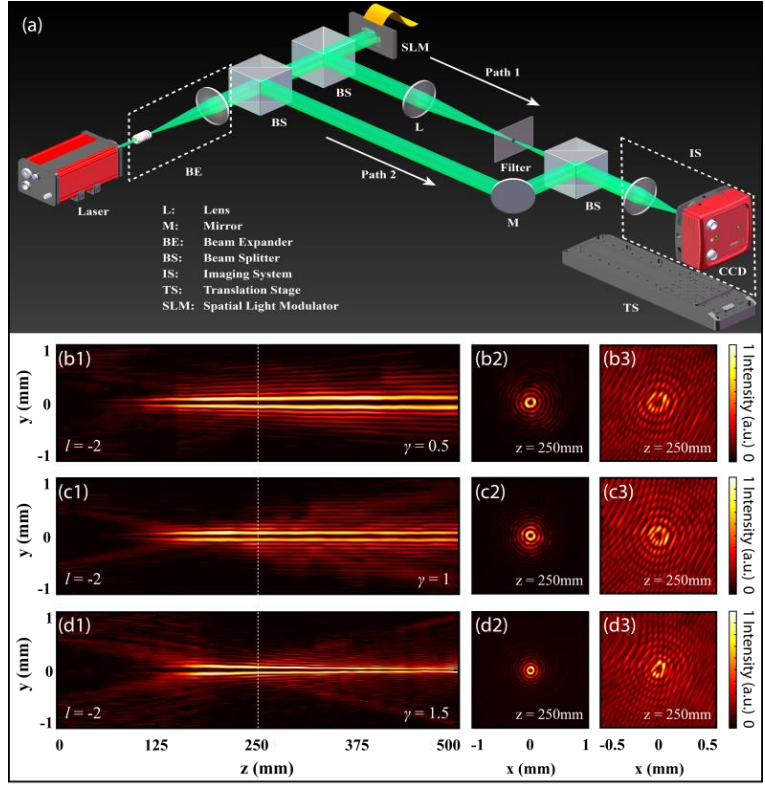

Fig. 1. Experimental observations of POVBs in free-space for different values of the phase modulation exponent $\gamma$. (a) Schematic of the experimental setup used for the generation and detection of the POVBs. Path 1 highlights the beamline used for the generation, while path 2 for carrying out the interferograms. (b1-d1) Normalized "sideview" of the beams in the $y$-$z$ plane for (b1) $\gamma$ = 0.5, (c1) $\gamma$ =1, and (d1) $\gamma$ =1.5. (b2-d2) Intensity distributions and (b3-d3) interferograms recorded at the distance $z$ = 250mm, marked by dashed white lines in (b1-d1). $l$ = -2 is the topological charge.

Figures 1(b1-d3) present experimental results associated with three different cases of POVBs, where the power-law phase exponent coefficient $\gamma$ takes the values 0.5, 1, and 1.5. Without loss of generality, the topological charge $l$ is chosen to be -2 for all the cases under consideration. The scaling parameter $C_\rho$ related to the three values of $\gamma$ is, respectively, 5.96, 3.77, and 3.12 µm, while $w_\rho$ is 1 mm. The choice of $C_\rho$ is not arbitrary, rather these coefficients are purposely calculated to obtain beam peak intensities appearing to a comparable propagation distance. The initial amplitude modulation is appropriately designed to generate POVBs exhibiting a low-varying peak intensity over a long range of distances. In this case, a quasi-constant peak intensity evolution can be achieved using the formula

$$A(\rho) = A_{\text{POVB}}(\rho) i^{|l|} \rho^{-\gamma/2} \sqrt{\frac{2-\gamma}{2\pi\gamma kC}}, \tag{4}$$

with $A_{\text{POVB}}(\rho)$ being a constant coefficient related to the average beam power. The initial Gaussian beam is significantly expanded (radius = 8 mm) to meet the theoretical plane wave approximation of the initial condition. However, in case the influence of the Gaussian beam truncation cannot be neglected, it can also be embedded in the amplitude modulation for improving the maximum propagation range. Looking at the longitudinal intensity evolution in Figs. 1(b1-d1), the POVBs undergo an initial autofocusing dynamic, reaching the highest peak intensity at a distance of about $z = 200$ mm, and then exhibiting a dramatic reshaping of the intensity pattern into a high-order Bessel-like beam. A better visualization is provided in Figs. 1(b2-d2), where the corresponding transverse intensity distributions recorded by the CCD camera at the distance $z = 250$ mm, are presented. For $\gamma = 0.5$, both the hollow-core radius and the annular main-lobe width of the Bessel-like beam increase during propagation. In contrast, an opposite propagation dynamic occurs as $\gamma$ takes a value larger than 1.0, say 1.5 as used in Fig. 1. Indeed, no appreciable variation of the beam size is observed for a unitary value $\gamma$ after the focal distance. Additionally, the beam intensity evolutions after the focal distance are examined. Since the initial amplitude is appropriately modulated to maintain a quasi-constant peak intensity during evolution, we observe a high intensity after 500 mm-long propagation in free space for all these three cases. For comparison, we also performed a series of measurements of the POVBs without including any initial amplitude. Under such conditions, the beam intensities reach a maximum at a certain distance and decay significantly afterward, with undetectable light at $z = 500$ mm, indicating the essential role played by amplitude modulation. Finally, interferograms are recorded to examine the topological charge carried by the POVBs (see Figs. 1(b3-d3)). For each case, the interference fringes display a fork structure from the phase singularity, indicating a topological charge $l$ of -2. These results clearly demonstrate the robustness of the OAM carried by the vortex pin beams.

### 4. Numerical Simulations

The above experimental observations can be corroborated with numerical simulations. To this purpose, Fig. 2 presents numerical results for the three cases corresponding to the measurements illustrated in Fig. 1. In particular, computations are carried out by solving Eq. (1) via the split-step Fourier transformed method. All values of the parameters used in simulations are consistent with physical dimensions in our experimental setting. For the sake of completeness, we also list the other parameters used in our simulations: $\rho_0 = 0.5$ mm, $\lambda = 532$ nm, $A_{\text{POVB}}(\rho) = 3.23$. The initial condition is constituted by a phase- and amplitude-modulated Gaussian beam, whose intensity value is zero inside a circle of radius $\rho_0$. This assumption is motivated by the fact that the amplitude is diverging at the axis origin. As shown in Figs. 2(a1-c1), the POVBs are observed to initially experience autofocusing dynamic at the earlier stage of their propagation followed by a subsequent reshaping into a high-order Bessel-like beam profile. Furthermore, normalized transverse intensity distributions retrieved at the distance $z = 250$ mm highlight the Bessel-like shape expected for the POVBs after the focal distance (Figs. 2(a2-c2)). For different values of the $\gamma$ coefficient, the beam main lobes behave differently during propagation, hence exhibiting a propagation dependence. This behavior is also expected from the analytical solution in Eq. (3). Indeed, if $R_{l0}$ is the full width at half-maximum (FWHM) of the $l$-order Bessel function in Eq. (2), the radius of a POVB is given by

$$R(z) = \frac{R_{l0}}{2k\left(C\gamma z^{\gamma-1}\right)^{\frac{1}{2-\gamma}}}, \tag{5}$$

which depends in general on the propagation distance $z$ except for the case $\gamma = 1$. $R(z)$ increases (or decreases) as $\gamma$ acquires a value smaller (or larger) than 1. Nevertheless, despite the progressive "expansion" or "shrinking" of the POVBs, their peak intensity remains almost constant over a long distance due to the initial amplitude modulation. Indeed, the peak intensity shows only slight oscillations around a constant value, with the oscillation frequency being increased for higher values of $\gamma$ (see solid white lines in Figs. 2(a1-c1)). It is worth pointing out again that amplitude modulation plays a primary role in the propagation dynamics of these vortex beams. Should only the phase modulation be applied to the initial Gaussian beam, the peak intensity of the POVB would have reached a maximum at a certain distance and then rapidly decayed to zero, similarly to the OPBs in Ref. [27]. The topological charges carried by the POVBs can be envisaged from their phase structures, as shown in Figs. 2(a3-c3). Wrapped phase patterns extracted at $z = 250$ mm from the numerical simulations show a clockwise spiral shape whose number of cycles is higher for low values of $\gamma$, and more importantly, they reveal a topological charge $l = -2$. Besides, in order to prove the validity of the analytical approximation on modelling the linear dynamics of POVBs, we also compare numerical results with analytical predictions from Eq. (3). Plots in Figs. 2(a4-c4) show the overlapped output intensity profiles after 500 mm-long propagation in free-space for all three cases of POVBs. The analytical curves match very well the numerical simulations, especially in proximity of the main-lobe ring. The agreement further improves for even longer distances, thus quantitatively confirming the reshaping of the amplitude- and phase-modulated Gaussian beam at the onset into a high-order Bessel-like shape in the far-field. These numerical and analytical results agree well with those from the experiments presented in Fig. 1.

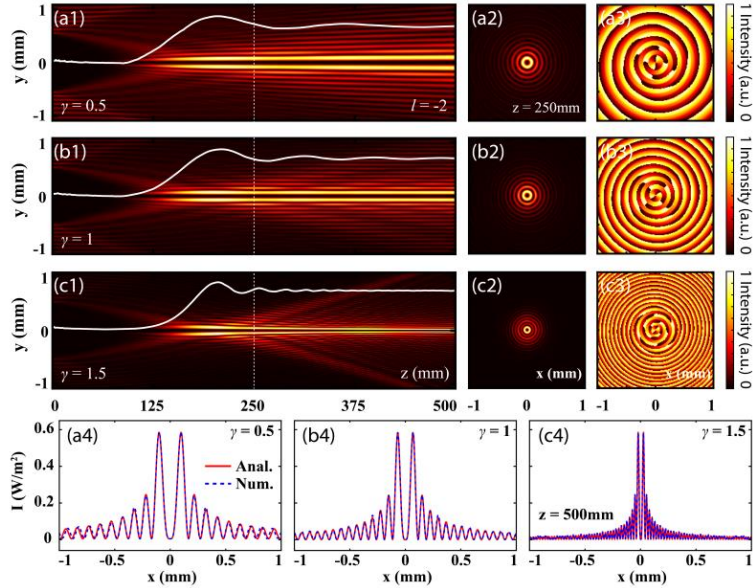

Fig. 2 Numerical simulation of POVBs for different values of the parameter $\gamma$, with topological charge $l = -2$ and initial amplitudes pre-designed to achieve low-varying peak intensity over a long range of distances. (a1-c1) Normalized longitudinal intensity distributions in the $y$-$z$ plane for (a1) $\gamma = 0.5$, (b1) 1, and (c1) 1.5, respectively; solid white lines in each panel plot the peak intensity evolutions. Corresponding (a2-c2) transverse intensity distributions and (a3-c3) wrapped phase patterns extracted at a distance $z = 250$mm, marked by dashed white lines in (a1-c1). (a4-c4) Comparison between numerical (blue dashed lines) intensity profiles in (a1-c1) and analytical (red solid lines) prediction from Eq. (3), performed at the output after 500mm-long propagation.

## 5. Control of the Peak Intensity Evolution

So far, the discussion has been only restricted to the PVOBs that display a constant or low-varying peak intensity evolution during subsequent propagation after initial autofocusing. Now we explore the tunability of the peak intensity over a long range of distances by properly engineering the initial amplitude modulation $A(\rho)$. Three typical examples are illustrated in Figs. 3(a1-c1), where the peak intensities follow a hyperbolic secant, a flat-top, and a sinusoidal curve (as outlined by the white solid lines in Fig. 3). For these three cases, the POVBs are generated with the same power-law phase exponent coefficient $\gamma = 1.5$, but under a different amplitude modulation as follows. The profile of the initial amplitude modulation $A(\rho)$ can be estimated by determining the function $A_{\text{POVB}}(\rho)$ to be inserted in Eq. (4) as:

| POVBs | Analytical peak intensity vs z |
|---|---|
| Steady-state | $I(z) = I_0$ with $I_0 = 2.467$ W/m$^2$ |
| Hyperbolic secant | $I(z) = I_0 \text{sech}(22z - 6.49)$ |
| Flat-top | $I(z) = I_0 \exp(-(7.106z - 2.096)^4)$ |
| Sinusoidal | $I(z) = I_0(1 - 0.5(\sin(6\pi z - 1.649))^2)$ |

Tab. 1 Analytical peak intensity of the POVBs in Fig. 3 as a function of $z$.

$$A_{\text{POVB}}(\rho) = 3.23\sqrt{I(z)\left(\rho^{2-\gamma}/(C\gamma)\right)/I_0}, \qquad (6)$$

where $I(z)$ are the analytical peak intensity curves as a function of $z$. The exact profiles of $A_{\text{POVB}}(\rho)$ for all specific cases of POVBs analyzed in this manuscript are listed in Tab. 1. Furthermore, we perform a series of experiments to demonstrate POVBs featuring such pre-designed peak intensity evolutions. Experimental results are presented in Figs. 3(a2-c2), showing a clear modulation of the maximum intensity value along the propagation direction. The measured intensity variations show an evolution trend similar to the predicted curves.

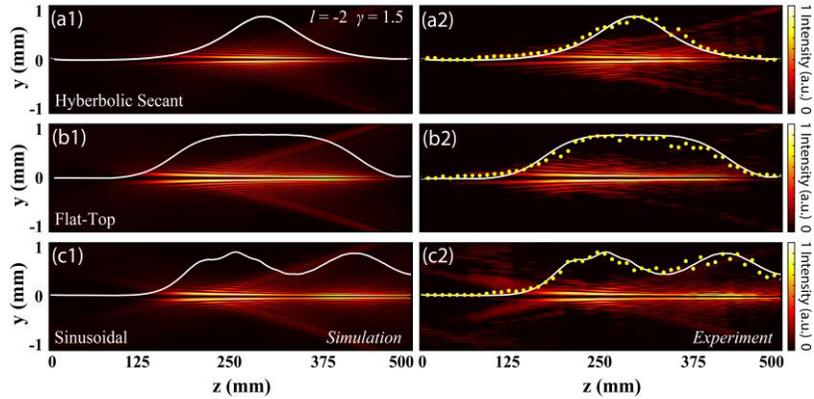

Fig. 3 Numerical (left) and experimental (right) results of POVBs in free-space with $\gamma = 1.5$, topological charge $l = -2$, and different initial amplitudes designed to achieve a desired peak intensity evolution. Left-column panels illustrate the normalized longitudinal intensity distributions of POVBs whose peak intensity along the propagation direction follows (a1) a hyperbolic secant, (b1) a flat-top, and (c1) a sinusoidal curve, as marked by a solid white line in each panel. Right-column panels show the corresponding experimental results. To provide a better comparison, experimental peak intensity evolutions (yellow dots) extracted from measurements at each distance $z$ are overlapped to the corresponding curves from predictions (solid white lines).

## 6. Poynting Vectors and Optical Forces

As seen above, the POVBs can be designed to exhibit tunable features, with either a constant or a pin-like vortex core as well as a controllable peak intensity evolution, in addition to preserved OAM during propagation. Besides the propagation properties, for both fundamental understanding and applications, it is also useful to investigate the energy-linked features associated with this class of optical beams. In this section, two key quantities related to the electromagnetic energy carried by POVBs are numerically explored: the Poynting vectors and the optical forces.

### 6.1 Poynting vectors of POVBs

The Poynting vector represents the energy flux of the radiation field per unit area and is defined as $\vec{S} = \mu_0^{-1} \vec{E} \times \vec{B}$, with $\vec{B} = \nabla \times \vec{U}$ and $\vec{E} = ic^2 \omega^{-1} \nabla \times \vec{B}$ denoting, respectively, the magnetic and electric field, while $\mu_0$ is the vacuum magnetic permeability, $c$ the speed of light, and $\nabla$ the nabla operator. For harmonic electromagnetic fields, the measured beam intensity directly relies on the time-averaged value of the Poynting vector over the wave cycle $T = 2\pi/\omega$. By applying the Lorentz gauge condition and the paraxial approximation, the average power flow can be expressed as [62, 63]

$$\langle \vec{S} \rangle = \frac{1}{2\mu_0}\left(\vec{E} \times \vec{B}^* + \vec{E}^* \times \vec{B}\right) = \frac{\omega}{2\mu_0}\left[i\left(\psi \nabla_\perp \psi^* - \psi^* \nabla_\perp \psi\right) + 2k|\psi|^2 \hat{z}\right], \quad (7)$$

where the symbol ($*$) denotes the complex conjugate, $\nabla_\perp = \partial/\partial x \,\hat{x} + \partial/\partial y \,\hat{y}$ is the transverse nabla operator, and $\hat{x}$, $\hat{y}$, and $\hat{z}$ are the unit vectors. In Eq. (7), the first-term on the right side refers to the transverse component of the Poynting vector in the $x$-$y$ plane, whereas the second-term describes the energy flux flowing along the propagation direction $z$, revealing a direct proportionality with the beam intensity.

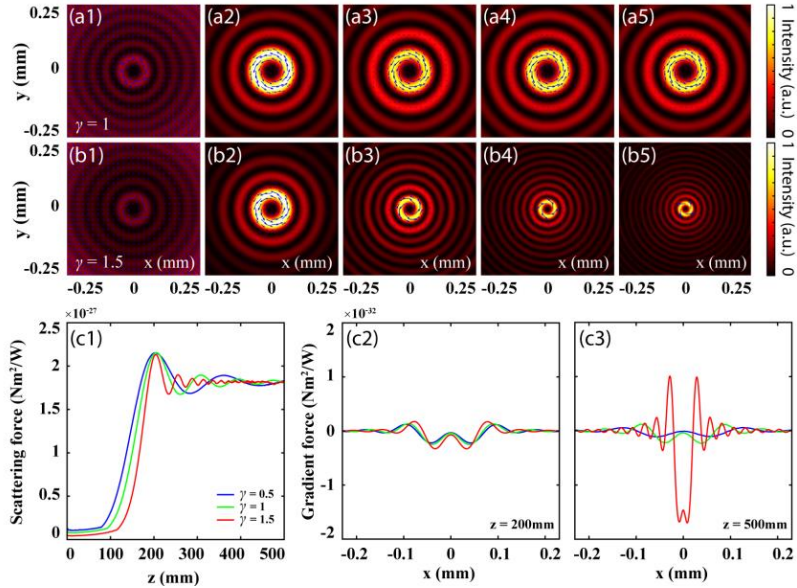

Fig. 4: Numerical calculations of Poynting vectors associated to POVBs with topological charge $l = -2$ for two different values of the exponent coefficient $\gamma$: (a1-a5) $\gamma = 1$, and (b1-b5) $\gamma = 1.5$. Background distributions show normalized intensity patterns of the POVBs at (varying) distances $z = 100, 200, 300, 400$, and 500mm. Blue arrows highlight the magnitude and direction of the transverse components of the Poynting vectors in the $x$-$y$ plane. (c1-c3) Numerical calculations of the optical forces on a test polystyrene sphere suspended in water exerted by the three POVBs studied in Fig. 1 and Fig. 2. (c1) Maximum scattering force as a function of the propagation distance $z$. (c2-c3) Transverse gradient force from the POVB, calculated at two selected distances (c2) $z = 200$, and (c3) 500mm.

Numerically-calculated Poynting vectors of the POVBs from Eq. (7) are illustrated in Figs. 4(a1-a5 and b1-b5) at selected distances for two different values of $\gamma$. In particular, blue arrows displaying the transverse components of the Poynting vectors are superimposed to the corresponding intensity distributions shown in the background. At the first stage, the energy flux radially flows from the beam sub-lobes towards the center (see Figs. 4(a1, b1)), leading to the focused vortex structure of the beam. After that, the energy flux mainly localizes and circulates around the high-intensity main lobe for subsequent long-distance propagation. Since for $\gamma = 1.5$ both the hollow-core radius and the main lobe width of the POVB decrease during propagation, the transverse energy flux around the main lobe becomes gradually more localized (Figs. 4(b1-b5)). In contrast, the power flux remains invariant for $\gamma = 1$ as shown in Figs. 4(a1-a5).

## 6.2 Optical forces of POVBs

Optical forces originate from the transfer of linear or angular momentum from an optical beam to an illuminated object, essential for optical trapping and manipulation. As an illustrative example, we study here the optical forces exerted by a linearly-polarized POVB on a spherical dielectric nanoparticle, suspended in a medium with a dielectric constant $\varepsilon_m$ and featuring a radius $a$ as well as a dielectric constant $\varepsilon_p$. If the particle size is sufficiently smaller than the light wavelength, the nanoparticle behaves like an electric dipole in accordance with Rayleigh scattering [64]. A simple description can be established by modelling the radiation pressure forces acting on an electromagnetic dipole, known as the Rayleigh regime, which is split into two components: the scattering and the gradient forces. The former is associated with the linear momentum changes through [64]

$$\vec{F}_{scat}(r,\theta,z) = \frac{n_m}{c} C_{pr} \langle \vec{S} \rangle_z,\tag{8}$$

where $\langle \vec{S} \rangle_z = I(r,\theta,z)\hat{z} = 0.5 n_m \varepsilon_0 c |\psi(r,\theta,z)|^2 \hat{z}$ is the longitudinal component of the time-averaged value of the Poynting vector, and $C_{pr}$ is the cross-section pressure of the particle. For isotropic particles, $C_{pr}$ is equal to the scattering cross-section $C_{scat}$:

$$C_{pr} = C_{scat} = \frac{8}{3}\pi k^4 a^6 \left(\frac{m^2-1}{m^2+2}\right)^2,\tag{9}$$

with $m = n_p / n_m$ being the ratio between the refractive index of the particle $n_p$ and that of the medium $n_m$. The other force, i.e., the gradient force, is related to the Lorentz force induced by the electromagnetic field on the dipole. In the steady-state, the time-averaged gradient force reduces to [64, 65]:

$$\vec{F}_{grad}(r,\theta,z) = \frac{2\pi n_m a^3}{c}\left(\frac{m^2-1}{m^2+2}\right)\nabla I(r,\theta,z).\tag{10}$$

Figures 4(c1-c3) present the scattering and gradient forces from the three POVBs studied earlier (Fig. 1 and Fig. 2), calculated numerically according to Eqs. (8) and (10). Specifically, we examined the case of polystyrene spheres (mass density 1050kg/m$^3$) suspended in water, with the parameters in the simulations chosen as $n_p=1.592$, $n_m=1.33$, and $a=20$nm. The scattering force, which acts mainly along the beam propagation direction, is shown in Fig. 4(c1), and the maximum scattering force exerted by the three POVBs follows the corresponding beam peak intensity evolution (see solid white lines in Fig. 2), reaching the highest value near the focal distance and then oscillating asymptotically for a long distance. On the other hand, the gradient force acts in the transverse direction and determines the ability of POVBs to trap a particle. Numerical results show that the gradient force associated with a higher value of $\gamma$ is much larger than for the other two cases with a lower value of $\gamma$, especially after a long propagation distance (see Figs. 4(c2 and c3)). Physically, the increased gradient force is mediated by the pin-like

feature of the vortex beams, as the hollow-core radius of the POVB gets closer to the size of the particle, thus guaranteeing a better trapping capability. We believe that our numerical study on the optical forces provides a guide for employing the POVBs in experiments of optical trapping applications, including three-dimensional particle manipulation [66].

## 7. Conclusion

In this work, we have experimentally demonstrated pin-like optical vortex beams and investigated their energy flow properties by numerically calculating the Poynting vectors and the optical radiation forces. These vortex beams are generated by modulating both the amplitude and the phase profile of an input laser beam through a spatial phase structure composed of a radially-symmetric power-law phase chirp profile and an orbital angular moment phase-term with an arbitrary topological charge. A POVB initially exhibits an autofocusing evolution and reaches the maximum intensity, then displays a reshaping of the amplitude pattern into a high-order Bessel-like beam, whose annular main-lobe width and hollow-core radius change during propagation. An appropriate design of the initial amplitude based on the theoretical analysis allows controlling the beam peak intensity along with the propagation range. Moreover, we have also shown that the transverse power flux initially flows radially from the beam sub-lobes towards the beam center to form the vortex structure, and then localizes and rotates with the higher values of the Poynting vectors concentrated in the annular beam main lobe. Finally, we have studied the optical radiation forces associated with these vortex pins and shown their dependence on the design parameters for the POVBs. These specially designed beams can lead to optimal optical forces for trapping and manipulating a nanoparticle under test. Moreover, this research expands the understanding of free-space generation of anti-diffracting optical vortex beams and builds up a connection between singular optics and structured light, which may find applications in different areas such as optical communication and quantum information technologies.

## Funding

We acknowledge financial support from the National Key R&D Program of China (2017YFA0303800), National Natural Science Foundation of China (NSFC) (61575098, 91750204, 11674180) and 111 Project in China (B07013), the NSERC Discovery and Strategic grants in Canada, and from the MESI in Quebec. D.B. acknowledges support from the 66 Postdoctoral Science Grant of China.

## Acknowledgments

We thank Ze Zhang for assistance. R.M. is affiliated to UESTC as an adjoining faculty.

## Disclosures

The authors declare no conflicts of interest.

## References

1. H. Rubinsztein-Dunlop, A. Forbes, M. V. Berry, M. R. Dennis, D. L. Andrews, M. Mansuripur, C. Denz, C. Alpmann, P. Banzer, T. Bauer, E. Karimi, L. Marrucci, M. Padgett, M. Ritsch-Marte, N. M. Litchinitser, N. P. Bigelow, C. Rosales-Guzmán, A. Belmonte, J. P. Torres, T. W. Neely, M. Baker, R. Gordon, A. B. Stilgoe, J. Romero, A. G. White, R. Fickler, A. E. Willner, G. Xie, B. McMorran, and A. M. Weiner, "Roadmap on structured light," J. Opt. **19**, 013001 (2017).
2. N. K. Efremidis, Z. Chen, M. Segev, and D. N. Christodoulides, "Airy beams and accelerating waves: an overview of recent advances," Optica **6**, 686-701 (2019).
3. M. S. Soskin and M. V. Vasnetsov, "Singular optics," Prog. Opt. **42**, 219-276 (2001).
4. A. S. Desyatnikov, L. Torner, and Y. S. Kivshar, "Optical vortices and vortex solitons," Prog. Opt. **47**, 291-391 (2005).
5. M. R. Dennis, K. O'Holleran, and M. J. Padgett, "Singular optics: optical vortices and polarization singularities," Prog. Opt. **53**, 293-363 (2009).
6. J. E. Curtis and D. G. Grier, "Structure of optical vortices," Phys. Rev. Lett. **90**, 133901 (2003).

7. Y. Shen, X. Wang, Z. Xie, C. Min, X. Fu, Q. Liu, M. Gong, and X. Yuan, "Optical vortices 30 years on: OAM manipulation from topological charge to multiple singularities," Light. Sci. Appl. **8**, 90 (2019).
8. N. B. Simpson, L. Allen, and M. J. Padgett, "Optical tweezers and optical spanners with Laguerre–Gaussian modes," J. Mod. Opt. **43**, 2485-2491 (1996).
9. H. He, M. E. Friese, N. R. Heckenberg, and H. Rubinsztein-Dunlop, "Direct observation of transfer of angular momentum to absorptive particles from a laser beam with a phase singularity," Phys. Rev. Lett. **75**, 826-829 (1995).
10. T. Stav, A. Faerman, E. Maguid, D. Oren, V. Kleiner, E. Hasman, and M. Segev, "Quantum entanglement of the spin and orbital angular momentum of photons using metamaterials," Science **361**, 1101-1104 (2018).
11. R. Fickler, G. Campbell, B. Buchler, P. K. Lam, and A. Zeilinger, "Quantum entanglement of angular momentum states with quantum numbers up to 10,010," PNAS **113**, 13642-13647 (2016).
12. K. T. Gahagan and G. A. Swartzlander, Jr., "Optical vortex trapping of particles," Opt. Lett. **21**, 827-829 (1996).
13. L. Allen, M. W. Beijersbergen, R. J. Spreeuw, and J. P. Woerdman, "Orbital angular momentum of light and the transformation of Laguerre-Gaussian laser modes," Phys. Rev. A **45**, 8185-8189 (1992).
14. F. Flossmann, U. T. Schwarz, and M. Maier, "Propagation dynamics of optical vortices in Laguerre–Gaussian beams," Opt. Commun. **250**, 218-230 (2005).
15. Y. Cai, X. Lu, and Q. Lin, "Hollow Gaussian beams and their propagation properties," Opt. Lett. **28**, 1084-1086 (2003).
16. J. Arlt and K. Dholakia, "Generation of high-order Bessel beams by use of an axicon," Opt. Commun. **177**, 297-301 (2000).
17. K. Volke-Sepulveda, V. Garcés-Chávez, S. Chávez-Cerda, J. Arlt, and K. Dholakia, "Orbital angular momentum of a high-order Bessel light beam," J. Opt. B Quantum Semiclassical Opt. **4**, S82-S89 (2002).
18. S. Chávez-Cerda, M. J. Padgett, I. Allison, G. H. C. New, J. C. Gutiérrez-Vega, A. T. O'Neil, I. MacVicar, and J. Courtial, "Holographic generation and orbital angular momentum of high-order Mathieu beams," J. Opt. B Quantum Semiclassical Opt. **4**, S52-S57 (2002).
19. C. Lopez-Mariscal, J. C. Gutierrez-Vega, G. Milne, and K. Dholakia, "Orbital angular momentum transfer in helical Mathieu beams," Opt. Express **14**, 4182-4187 (2006).
20. Y. F. Chen, T. H. Lu, and K. F. Huang, "Observation of spatially coherent polarization vector fields and visualization of vector singularities," Phys. Rev. Lett. **96**, 033901 (2006).
21. C. Rosales-Guzmán, B. Ndagano, and A. Forbes, "A review of complex vector light fields and their applications," J. Opt. **20**, 123001 (2018).
22. J. Arlt, T. Hitomi, and K. Dholakia, "Atom guiding along Laguerre-Gaussian and Bessel light beams," Appl. Phys. B **71**, 549-554 (2000).
23. M. Goutsoulas, D. Bongiovanni, D. Li, Z. Chen, and N. K. Efremidis, "Tunable self-similar Bessel-like beams of arbitrary order," Opt. Lett. **45**, 1830-1833 (2020).
24. R. Grunwald and M. Bock, "Needle beams: a review," Adv. Phys. X **5**, 1736950 (2020).
25. X. Weng, Q. Song, X. Li, X. Gao, H. Guo, J. Qu, and S. Zhuang, "Free-space creation of ultralong anti-diffracting beam with multiple energy oscillations adjusted using optical pen," Nat. Commun. **9**, 5035 (2018).
26. C. Vetter, R. Steinkopf, K. Bergner, M. Ornigotti, S. Nolte, H. Gross, and A. Szameit, "Realization of free-space long-distance self-healing Bessel beams," Laser Photon. Rev. **13**, 1900103 (2019).
27. Z. Zhang, X. Liang, M. Goutsoulas, D. Li, X. Yang, S. Yin, J. Xu, D. N. Christodoulides, N. K. Efremidis, and Z. Chen, "Robust propagation of pin-like optical beam through atmospheric turbulence," APL Photon. **4**, 076103 (2019).
28. Z. Chen, M. Segev, and D. N. Christodoulides, "Optical spatial solitons: historical overview and recent advances," Rep. Prog. Phys. **75**, 086401 (2012).
29. Y. V. Kartashov, B. A. Malomed, and L. Torner, "Solitons in nonlinear lattices," Rev. Mod. Phys. **83**, 247-305 (2011).
30. R. Gautam, A. Bezryadina, Y. Xiang, T. Hansson, Y. Liang, G. Liang, J. Lamstein, N. Perez, B. Wetzel, R. Morandotti, and Z. Chen, "Nonlinear optical response and self-trapping of light in biological suspensions," Adv. Phys. X **5**, 1778526 (2020).
31. L. Tang, D. Song, S. Xia, S. Ma, W. Yan, Y. Hu, J. Xu, D. Leykam, and Z. Chen, "Photonic flat-band lattices and unconventional light localization," Nanophotonics **9**, 1161-1176 (2020).
32. J. Durnin, J. J. Miceli, and J. H. Eberly, "Diffraction-free beams," Phys. Rev. Lett. **58**, 1499-1501 (1987).
33. J. Durnin, "Exact solutions for nondiffracting beams I The scalar theory," J. Opt. Soc. Am. A **4**, 651-654 (1987).
34. J. C. Gutierrez-Vega, M. D. Iturbe-Castillo, and S. Chavez-Cerda, "Alternative formulation for invariant optical fields: Mathieu beams," Opt. Lett. **25**, 1493-1495 (2000).
35. P. Zhang, Y. Hu, T. Li, D. Cannan, X. Yin, R. Morandotti, Z. Chen, and X. Zhang, "Nonparaxial Mathieu and Weber accelerating beams," Phys. Rev. Lett. **109**, 193901 (2012).
36. I. D. Chremmos, Z. Chen, D. N. Christodoulides, and N. K. Efremidis, "Bessel-like optical beams with arbitrary trajectories," Opt. Lett. **37**, 5003-5005 (2012).
37. J. Zhao, P. Zhang, D. Deng, J. Liu, Y. Gao, I. D. Chremmos, N. K. Efremidis, D. N. Christodoulides, and Z. Chen, "Observation of self-accelerating Bessel-like optical beams along arbitrary trajectories," Opt. Lett. **38**, 498-500 (2013).
38. M. Goutsoulas and N. K. Efremidis, "Precise amplitude, trajectory, and beam-width control of accelerating and abruptly autofocusing beams," Phys. Rev. A **97**, 063831 (2018).
39. G. A. Siviloglou and D. N. Christodoulides, "Accelerating finite energy Airy beams," Opt. Lett. **32**, 979-981 (2007).

40. G. A. Siviloglou, J. Broky, A. Dogariu, and D. N. Christodoulides, "Observation of accelerating Airy beams," Phys. Rev. Lett. **99**, 213901 (2007).
41. M. A. Bandres, "Accelerating beams," Opt. Lett. **34**, 3791-3793 (2009).
42. M. A. Bandres, "Accelerating parabolic beams," Opt. Lett. **33**, 1678-1680 (2008).
43. N. K. Efremidis and D. N. Christodoulides, "Abruptly autofocusing waves," Opt. Lett. **35**, 4045-4047 (2010).
44. P. Zhang, J. Prakash, Z. Zhang, M. S. Mills, N. K. Efremidis, D. N. Christodoulides, and Z. Chen, "Trapping and guiding microparticles with morphing autofocusing Airy beams," Opt. Lett. **36**, 2883-2885 (2011).
45. D. G. Papazoglou, N. K. Efremidis, D. N. Christodoulides, and S. Tzortzakis, "Observation of abruptly autofocusing waves," Opt. Lett. **36**, 1842-1844 (2011).
46. I. Chremmos, N. K. Efremidis, and D. N. Christodoulides, "Pre-engineered abruptly autofocusing beams," Opt. Lett. **36**, 1890-1892 (2011).
47. S. N. Khonina, A. P. Porfirev, and A. V. Ustinov, "Sudden autofocusing of superlinear chirp beams," J. Opt. **20**, 025605 (2018).
48. B. Chen, C. Chen, X. Peng, Y. Peng, M. Zhou, and D. Deng, "Propagation of sharply autofocused ring Airy Gaussian vortex beams," Opt. Express **23**, 19288-19298 (2015).
49. F. O. Fahrbach, P. Simon, and A. Rohrbach, "Microscopy with self-reconstructing beams," Nat. Photon. **4**, 780-785 (2010).
50. T. Vettenburg, H. I. Dalgarno, J. Nylk, C. Coll-Llado, D. E. Ferrier, T. Cizmar, F. J. Gunn-Moore, and K. Dholakia, "Light-sheet microscopy using an Airy beam," Nat. Methods **11**, 541-544 (2014).
51. S. Jia, J. C. Vaughan, and X. Zhuang, "Isotropic 3D super-resolution imaging with a self-bending point spread function," Nat. Photon. **8**, 302-306 (2014).
52. P. Polynkin, M. Kolesik, J. V. Moloney, G. A. Siviloglou, and D. N. Christodoulides, "Curved plasma channel generation using ultraintense Airy beams," Science **324**, 229-232 (2009).
53. J. Baumgartl, M. Mazilu, and K. Dholakia, "Optically mediated particle clearing using Airy wavepackets," Nat. Photon. **2**, 675-678 (2008).
54. S. Chen, S. Li, Y. Zhao, J. Liu, L. Zhu, A. Wang, J. Du, L. Shen, and J. Wang, "Demonstration of 20-Gbit/s high-speed Bessel beam encoding/decoding link with adaptive turbulence compensation," Opt. Lett. **41**, 4680-4683 (2016).
55. N. Ahmed, Z. Zhao, L. Li, H. Huang, M. P. Lavery, P. Liao, Y. Yan, Z. Wang, G. Xie, Y. Ren, A. Almaiman, A. J. Willner, S. Ashrafi, A. F. Molisch, M. Tur, and A. E. Willner, "Mode-division-multiplexing of multiple Bessel-Gaussian beams carrying orbital-angular-momentum for obstruction-tolerant free-space optical and millimetre-wave communication Links," Sci. Rep. **6**, 22082 (2016).
56. E. DelRe, A. J. Agranat, and C. Conti, "Light with no spatial scale: diffraction cancellation, anti-diffraction, scale-free instability and subwavelength beam propagation in dipolar glasses," in *Adv. Phot. Cong.*, OSA Technical Digest (online) (Optical Society of America, 2012), NW3D.1.
57. E. DelRe, E. Spinozzi, A. J. Agranat, and C. Conti, "Scale-free optics and diffractionless waves in nanodisordered ferroelectrics," Nat. Photon. **5**, 39-42 (2010).
58. E. DelRe, F. Di Mei, J. Parravicini, G. Parravicini, A. J. Agranat, and C. Conti, "Subwavelength anti-diffracting beams propagating over more than 1,000 Rayleigh lengths," Nat. Photon. **9**, 228-232 (2015).
59. H. Harutyunyan, "Anti-diffraction of light," Nat. Photon. **9**, 213-214 (2015).
60. D. Li, D. Bongiovanni, M. Goutsoulas, S. Xia, Z. Zhang, Y. Hu, D. Song, R. Morandotti, N. K. Efremidis, and Z. Chen, "Direct comparison of anti-diffracting optical pin beams and abruptly autofocusing beams," OSA Continuum **3**, 1525-1535 (2020).
61. J. A. Davis, D. M. Cottrell, J. Campos, M. J. Yzuel, and I. Moreno, "Encoding amplitude information onto phase-only filters," Appl. Opt. **38**, 5004-5013 (1999).
62. L. Allen, M. J. Padgett, and M. Babiker, "The orbital angular momentum of light," Prog. Opt. **39**, 291-372 (1999).
63. G. Chen, X. Huang, C. Xu, L. Huang, J. Xie, and D. Deng, "Propagation properties of autofocusing off-axis hollow vortex Gaussian beams in free space," Opt. Express **27**, 6357-6369 (2019).
64. Y. Harada and T. Asakura, "Radiation forces on a dielectric sphere in the Rayleigh scattering regime," Opt. Commun. **124**, 529-541 (1996).
65. Y. Roichman, B. Sun, Y. Roichman, J. Amato-Grill, and D. G. Grier, "Optical forces arising from phase gradients," Phys. Rev. Lett. **100**, 013602 (2008).
66. J. Zhao, I. D. Chremmos, D. Song, D. N. Christodoulides, N. K. Efremidis, and Z. Chen, "Curved singular beams for three-dimensional particle manipulation," Sci. Rep. **5**, 12086 (2015).